\documentclass[aps,prd,floatfix,twocolumn,footinbib,altaffilletter]{revtex4}

\usepackage[colorlinks=true,citecolor=blue,linkcolor=blue]{hyperref}
\usepackage{amsmath,amssymb}
\usepackage{epsfig}  
\usepackage{graphicx}               % Standard graphics package  
\usepackage{url}
\usepackage{hyperref}
\usepackage{color}
\usepackage{multirow}
\usepackage{placeins}

\usepackage{tikz}
\usetikzlibrary{trees}
\usetikzlibrary{decorations.pathmorphing}
\usetikzlibrary{decorations.markings}

  \definecolor{jblue}  {RGB}{20,50,100}
  \definecolor{npurple}  {RGB} {153, 51, 204}
  \definecolor{wred}   {RGB}{217,0,56}
  \definecolor{white}   {RGB}{255,255,255}
  
  \definecolor{korange}   {RGB}{235, 80,  43}
  \definecolor{korange2}   {RGB}{245, 100,  63}
  \definecolor{kyelloworange}   {RGB}{255, 210,  110}
  \definecolor{kyelloworange2}   {RGB}{240, 170,  90}
  \definecolor{kred}   {RGB}{204,  102, 153}
  \definecolor{kpurple}   {RGB}{153,  61, 190}
  \definecolor{kpurplelight}   {RGB}{213,  161, 230}

\tikzset{
	    gaugeboson/.style={decorate,decoration={snake, amplitude = 6pt, post length = 1 pt, pre length = 4 pt},draw=magenta},
	    fermion/.style={draw=black,postaction={decorate},decoration={markings,mark=at position .55}},
	    fermionin/.style={draw=black,postaction={decorate},decoration={markings,mark=at position .55 with {\arrow[draw=black]{<}}}},
	    fermionout/.style={draw=black,postaction={decorate},decoration={markings,mark=at position .55 with {\arrow[draw=black]{>}}}},
	    gluon/.style={decorate,draw=magenta,decoration={coil,amplitude = 6pt,segment length=8pt}},
	    connect/.style={draw=black,postaction={decorate},decoration={markings}}, %,mark=at position .55 with {\arrow[draw=red]{<}}}}
	    gluon2/.style={decorate,draw=magenta,decoration={coil,amplitude = 3pt,segment length=4pt}},
	    gaugeboson2/.style={decorate,decoration={snake, amplitude = 3pt, segment length=6pt},draw=black}
	}
	
\tikzset{
	  photon/.style={decorate, decoration={snake}, draw=npurple,very thick},
	  boson/.style={decorate, decoration={snake}, draw=npurple,very thick},
	  electron/.style={draw=jblue,very thick, postaction={decorate},
	           decoration={markings,mark=at position .55 with {\arrow[draw=jblue]{>}}}
	  },
	  electron2/.style={draw=jblue,very thick, postaction={decorate},
	           decoration={markings,mark=at position .55 with {\arrow[draw=jblue]{<}}}
	  },
	  fermion/.style={draw=jblue,very thick, postaction={decorate},
	            decoration={markings,mark=at position .55 with {\arrow[draw=jblue]{}}}
	  },
	  gluon/.style={decorate, draw=korange,very thick, %kred
	    decoration={coil,amplitude=4pt, segment length=6pt}},
	  higgs/.style={draw=wred,very thick, postaction={decorate},
	           decoration={markings,mark=at position .55 with {\arrow[draw=wred]{>}}}
	  },
	  nothing/.style={draw=white,very thick}
	}

\begin{document}

\title{The High-Energy Behavior of Photon, Neutrino and Proton Cross Sections}

\author{Carlos A. Arg\"uelles\footnote{arguelles@wisc.edu}}
\author{Francis Halzen\footnote{francis.halzen@icecube.wisc.edu}}
\author{Logan Wille\footnote{wille3@wisc.edu}}
\affiliation{Department of Physics, University of Wisconsin, Madison, WI 53706, USA} 
\affiliation{Wisconsin IceCube Particle Astrophysics Center, Madison, WI 53703, USA}
\author{Mike Kroll\footnote{mike.kroll@icecube.wisc.edu}}
\affiliation{Department of Physics, TU Dortmund University, D-44221 Dortmund, Germany}
\author{Mary Hall Reno\footnote{mary-hall-reno@uiowa.edu}}
\affiliation{Department of Physics and Astronomy, University of Iowa, Iowa City, Iowa 52242}

\begin{abstract}
By combining the color dipole model of the nucleon with the assumption that cross sections behave asymptotically as $\ln^2(s)$, we are able to describe the data for photon, neutrino and hadron interactions with protons at all energies; $s$ is the center-of-mass energy of the interacting particles. Specifically, we extrapolate the perturbative QCD calculations into the regime of small fractional parton momenta $x$ using a color dipole description of the proton target that guarantees an asymptotic $\ln^2(s)$ behavior of all cross sections. The ambiguity of introducing a parametrization associated with the dipole approximation is mitigated by the requirement that the saturation of the small-$x$ structure functions produces $\ln^2(s)$-behaved asymptotic cross sections, in agreement with the data. The same formalism allows us to calculate the cross section for the hadronic pair production of charm particles. The results, in particular those for the high-energy neutrino and charm cross sections, are relevant for evaluating the sensitivity as well as the background in neutrino telescopes.
\end{abstract}

\maketitle

\section{Introduction}
\label{sec:intro}

The high-energy behavior of photon, neutrino and proton cross sections on protons cannot be calculated perturbatively when the fractional momenta $x$ carried by the constituents become vanishingly small \cite{Altarelli:1977zs,Dokshitzer:1977,Gribov:1972}. The structure functions develop a $\ln ({1}/{x}$) divergent behavior that results in a violation of unitarity bounds \cite{Froissart:1961rq}.
It has been argued for some time that accelerator and cosmic ray data favor a $\ln^2(s)$ behavior of hadronic cross sections \cite{Block:2006hy}. In fact, a model-independent analytic extrapolation of a $\ln^2(s)$ description of the lower energy data on proton-proton total cross sections correctly anticipated \cite{Block:2011ak,Block:xq,Block:2012nj} the measurements at the LHC and the Auger cosmic ray observatory \cite{Auger:2013fv,Block:1999ss,Block:2000hs}. In this paper, we extend this successful phenomenological approach to photon and neutrino cross sections.

We present the unified dipole model framework \cite{Nikolaev:1990ja,Nikolaev:1993th,Mueller:1993rr,
Mueller:1994jq,Golec-Biernat:zk,Iancu:2003ge} that describes the behavior of $\gamma p$, $\nu p$ and $pp$ cross sections at high energies and small-$x$.  
Perturbative QCD calculations break down at high energy when the proton has an increasing number of partons with small fractional momenta $x$. 
In a parton picture one can simply think of saturation as screening resulting from the fact that the increasing number of small-$x$ partons have to be confined to a high energy proton of finite size.
Asymptotically, the proton is a black disk of (mostly) gluons with a radius that increases as $\ln s$.
%To remedy the unphysical small-$x$ behavior, a dipole parametrization of the proton target allows one to %extrapolate the perturbative calculations to low-$x$ values and high energy %\cite{Nikolaev:1990ja,Nikolaev:1993th,Mueller:1993rr,
%Mueller:1994jq,Golec-Biernat:zk,Iancu:2003ge}. 
The dipole description of the proton incorporates color saturation \cite{Golec-Biernat:zk}. 
What is new here is that we constrain the parameterization associated with the dipole model by the requirement that the cross section behaves asymptotically as $\ln^2(s)$ \cite{Block:2011ak,Block:2012nj}. 
%The ambiguity of introducing a parametrization associated with the dipole approximation is mitigated by the %requirement that the saturation of the small-$x$ structure functions produces a $\ln^2(s)$-behaved %asymptotic cross sections, in agreement with the data. 
Our main result is that a single dipole parameterization that incorporates saturation in this way 
%that yields an asymptotic $\ln^2(s)$ behavior of the cross section 
results in a successful description of all high-energy cross section data.

 %In this paper, saturation results in $\ln^2(s)$-behaved asymptotic cross sections. 

Our results are relevant to high-energy neutrino detectors \cite{IC:2013bs} whose sensitivity is directly proportional to the neutrino cross section. When operating in the PeV regime, where IceCube recently discovered a flux of cosmic neutrinos, the neutrino cross section can be calculated perturbatively with an accuracy of better than 5\%, constrained by measured HERA structure functions \cite{Gandhi:1998im, Choudhury:2011la,Connolly:2011qy,Cooper-Sarkar:2011fc,Henley:2006lp}. At EeV energies, relevant for the detection of cosmogenic neutrinos produced in the interactions of cosmic rays with background microwave photons, this is no longer the case and saturation effects must be included in evaluating the sensitivity of ARA \cite{Allison:2014kha}, ANITA\cite{ANITA:2010kq}, ARIANNA \cite{Gerhardt:2010fp}, JEM-EUSO \cite{JEMEUSO:2011aa}, and LUNASKA \cite{Bray:2013pi}. Our formalism provides a prediction of the EeV-neutrino cross section that, although relying on a (dipole) parameterization, is directly supported by a wealth of data.

The same framework can be used to predict the hadronic production of charm particles, for instance by
cosmic ray-air interactions 
in the atmosphere \cite{Gondolo:1995fq,Pasquali:1998ji,Gaisser:2013rx,Gelmini:2000wm,Martin:2003bu}. Above energies of $\sim 100$\,TeV, charmed hadron decay into neutrinos is the dominant atmospheric background for the detection of cosmic neutrinos. 
%The reason is that charm particles decay promptly into neutrinos, unlike long-lived high-energy pions and %kaons that, at these high energies, interact and lose energy before decaying. 
While perturbative QCD calculations are less reliable at high energies since saturation effects, associated with $m_c / \sqrt{s} \ll1$, are not included, our scheme includes these effects and makes definite predictions for the cross section using a dipole form that incorporates a wide range of data.
%%HR further constraining our result by fitting a wide range of data. 
%In the end, the charm production
We find that our charm production
cross section is in good agreement with previous studies using other dipole model parameterizations ~\cite{Enberg:2008fb,Goncalves:2006ch,Goncalves:2009qr}. 
%Although neutrinos of charm origin represent the dominant background at the highest energies, they have %not been observed because the data are dominated by conventional atmospheric neutrinos at low energy and %by cosmic neutrinos at high energy; charm neutrinos never dominate the measured spectrum. 
The charm production cross section and charm energy distribution is of great interest because a poor understanding of the charm neutrino background interferes with the precise characterization of the cosmic flux observed by IceCube \cite{Aartsen:2014kq}.

The paper is organized as follows: in Section \ref{sec:gamma-proton-xs} we review the $\gamma^* p$ cross sections in the dipole formalism; 
in Section \ref{sec:neutrino-proton-xs} we use a hybrid pQCD and dipole approach to calculate the $\nu N$ cross section; in Section \ref{sec:proton-proton-xs} we calculate the $pp \to c\bar{c} X$ and total $pp$ cross sections in the dipole formalism. We conclude in Section \ref{sec:conclu}.

\section{The Dipole Framework for Photon-Proton Cross Sections}
\label{sec:gamma-proton-xs} 

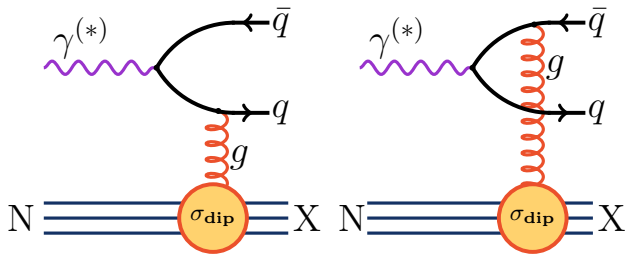
\begin{figure}
	\begin{tikzpicture}[line width=1.5 pt, scale=1.]
		\draw[photon] (0.5,0) -- (2, 0);
		\draw[gluon] (2.8,-0.57) -- (2.8,-1.8);
		
		\draw[fermionin] (3.0,0.6 ) -- (3.5,0.6);
		\draw[connect] (2,0) arc (150:90:1.2);
		
		\draw[fermionout] (3.0,-0.6 ) -- (3.5,-0.6);
		\draw[connect] (2,0) arc (210:270:1.2);
		
		\node at (1, 0.4) {\Large $\gamma^{(*)}$};
		\node at (3.1, -1.2) {\Large $g$};
		\node at (3.65, 0.6) {\Large $\bar{q}$};
		\node at (3.65, -0.6) {\Large $q$};
		
		\node at (0.2, -2.00) {\Large $\mathrm{N}$};
		\node at (4.0, -2.00) {\Large $\mathrm{X}$};
		\draw[fermion] (0.5,-1.8 ) -- (3.75,-1.8);
		\draw[fermion] (0.5,-2.0 ) -- (3.75, -2.0);
		\draw[fermion] (0.5,-2.2 ) -- (3.75,-2.2);
		
		\fill[kyelloworange]   (2.75,-2)  circle (0.45cm);
		\draw[korange]         (2.75,-2)  circle (0.45cm);
		\node[draw=none,fill=none] at (2.75,-2){$\bf \sigma_{dip}$}; 
		
		\fill[black] (2,0) circle (.04cm);
		\fill[black] (2.82,-0.575) circle (.04cm);
		% % % % % % % % % % % % % % % % % % % % % % % %
		\draw[photon] (4.7,0) -- (6.2, 0);
		\draw[gluon] 	  (7,0.57) -- (7,-1.8);
		
		%\draw[connect] (3.8, 0.6) -- (7.3, 0.6);
		%\draw[connect] (3.8, -0.6) -- (7.3, -0.6);
		
		\draw[fermionin]  (7.2,0.6 ) -- (7.7,0.6);
		\draw[connect]  (6.2,0) arc (150:90:1.2);
		
		\draw[fermionout] (7.2,-0.6 ) -- (7.7,-0.6);
		\draw[connect] (6.2,0) arc (210:270:1.2);
		
		\node at          (5.2, 0.4) {\Large $\gamma^{(*)}$};
		\node at 		  (7.3, -0) {\Large $g$};
		\node at 		  (7.85, 0.6) {\Large $\bar{q}$};
		\node at 		  (7.85, -0.6) {\Large $q$};
		
		\fill[black]      (6.2,0) circle (.04cm);
		\fill[black]      (7.02,0.575) circle (.04cm);
		
%		\draw[connect] (8.1 ,  0.6) -- (8.2,  0.6);
%		\draw[connect] (8.1 , -0.6) -- (8.2, -0.6);
%		\draw[connect] (8.15, -0.6) -- (8.15,-0.15);
%		\draw[connect] (8.15,  0.6) -- (8.15, 0.15);
%		\node at (8.15, 0.0) {\large $\boldsymbol{r}_\perp$};
		
		\node at (4.6, -2.00) {\Large $\mathrm{N}$};
		\node at (8.05, -2.00) {\Large $\mathrm{X}$};
		\draw[fermion] (4.8,-1.8 ) -- (7.8,-1.8);
		\draw[fermion] (4.8,-2.0 ) -- (7.8,-2.0);
		\draw[fermion] (4.8,-2.2 ) -- (7.8,-2.2);
		
		%pdf circle
		\fill[kyelloworange]   (7,-2)  circle (0.45cm);
		\draw[korange]         (7,-2)  circle (0.45cm);
		\node[draw=none,fill=none] at (7,-2){$\bf \sigma_{dip}$}; 
	
	\end{tikzpicture}
\caption{Diagram for the $\gamma^{(*)} p \to q \bar{q} X$ interaction in the dipole picture. }
\label{fig:diagram-gamma}
\end{figure}

In the color dipole picture of the proton  \cite{Nikolaev:1990ja,Nikolaev:1993th,Mueller:1993rr,
Mueller:1994jq,Golec-Biernat:zk,Iancu:2003ge,Ewerz:fp,Ewerz:ve}, the virtual photon-proton production cross section for producing $q\bar{q}$ pairs, shown in Figure \ref{fig:diagram-gamma}, has longitudinal (L) and transverse (T)  contributions given by
%
%The photon-proton $q\bar{q}$ production cross section for an incoming high energy photon, $ E_\gamma $ (i.e. $s \approx 2 m_p E_\gamma$), in the dipole picture is given by
%
\begin{equation}
\label{eq:ph-p-xs}
\sigma_{\gamma^* p \to q\bar{q} + X}^{T/L}(E_\gamma) = \int dz\ d^2 {\bf r} \ | \psi^\gamma_{T/L,q} (z,{\bf r};Q^2)|^2 \sigma_{dip} (x,{\bf r}) , ~ 
\end{equation}
where $ E_\gamma $ is the energy of the photon and $x$ the fractional momentum carried by the quark of the struck proton. The quantity $| \psi^\gamma_{T/L,q} (z,{\bf r};Q^2)|^2$ is the probability that $\gamma^*$ fluctuates into a $q \bar{q}$ pair with
 transverse separation $r$ and longitudinal momentum fraction $z$  and is given by \cite{Nikolaev:1990ja,Soyez:2007kg}
%
%where $x = (2 m_q)^2/s$, $| \psi^\gamma_{T,q} (z,{\bf r};Q^2)|^2$ is the probability that $\gamma$ fluctuates into a $q \bar{q}$ pair with separation ${\bf r}$ and transverse momentum fraction $z$ given by \cite{Nikolaev:1990ja,Enberg:2008fb}
%
\begin{equation}
\label{eq:ph-wavefunction-T}
| \psi^\gamma_{T,q}|^2 = e^2_q N_c  \frac{ \alpha_{em}}{2 \pi^2} \left[ (z^2 + \bar{z}^2) \epsilon^2 K^2_1(\epsilon r) + m^2_q K^2_0(\epsilon r) \right]
\end{equation}
\begin{equation}
\label{eq:ph-wavefunction-L}
| \psi^\gamma_{L,q}|^2 = e^2_q N_c  \frac{ \alpha_{em}}{2 \pi^2} (4 Q^2  z^2 \bar{z}^2) K^2_0(\epsilon r) 
\end{equation}
where $e_q$ is the quark charge, $N_c$ the number of colors, $\epsilon^2 = z \bar{z} Q^2 + m^2_q$, $\bar{z} = 1-z$, and $K_0$ and $K_1$ the modified Bessel functions. 

We have extracted $\sigma_{dip}$ from the deep inelastic structure function $F^{\gamma^* p}_2(x,Q^2)$ (see, for example, \cite{Ewerz:2012qq,Ewerz:ph}) using the high $Q^2$ approximation \cite{Ewerz:2011jt} 
\begin{equation}
\sigma_{dip}(x,r) = \pi^3 r^2 Q^2 \frac{\partial}{\partial Q^2} F^{\gamma p}_{2} |_{Q^2 =(z_0/r)^2}.
\label{eq:dip-F2-rel}
%HR: put in factor of Q2
\end{equation}
with $z_0 = 2.4$. Exploiting the convenient parametrization of $F_2(x,Q^2)$ by Block et al. \cite{Block:2013rr,Berger:gb}, using
their most recent result \cite{Block:2014fk}
for small $x$,  we obtain:
\begin{eqnarray}
\label{eq:dip-mad}
    \sigma_{dip}(x,r) &= &d_0  \frac{\pi ^3 r^2 (1-x)^n}{\tilde{r}^2+z_0^2} \bigg[ a_0 \tilde{r}^2 + c_1
   z_0^2 \nonumber \\
    &+& \mathcal{A} \left(a_1
   z_0^2+2 \mathcal{B} \left(a_2
   z_0^2+b_1 \tilde{r}^2+b_2 \tilde{r}^2
   \mathcal{B} \right) + 2 b_0 \tilde{r}^2 \right) \nonumber \\
   &+& \tilde{r}^2 \mathcal{B}
   \left(a_1+a_2 \mathcal{B} \right)+z_0^2 \mathcal{A}^2
   \left(b_1+2 b_2
   \mathcal{B} \right)\bigg] \\
   \mathcal{A} &=& \log\left(\frac{z_0^2}{\tilde{r}^2
   x+z_0^2 x}\right) \\
   \mathcal{B} &=& \log \left( 1 + \frac{z_0^2}{\tilde{r}^2} \right)
\end{eqnarray}
where $\tilde{r} = \mu r$ and the dipole parameters are given in Table \ref{tbl:model-param}. Our approach is to include primarily the
small-$x$ behavior so an overall normalization
factor $d_0$ is also required \cite{Jeong:2014mla}.
\begin{center}
\begin{table}
    \begin{tabular}{| l || c | }
    \hline
    Parameter & Value \\ \hline \hline
    $c_1$ & $1.475\times 10^{-1}$  \\ \hline
    $a_0$ & $8.205 \times 10^{-4}$  \\ \hline
    $a_1$ & $-5.148 \times 10^{-2}$  \\ \hline
    $a_2$ & $-4.725 \times 10^{-3}$ \\ \hline
    $b_0$ & $2.217 \times 10^{-3}$  \\ \hline
    $b_1$ & $1.244 \times 10^{-2}$  \\ \hline
    $b_2$ & $5.958 \times 10^{-4}$  \\ \hline
    $d_0$ & $0.71$  \\ \hline
    $\mu^2$ & $2.82\ {\rm GeV^2}$  \\ \hline
    $z_0$ & $2.4$  \\ \hline
    \hline 
  \end{tabular}
\caption{Values of our model parameters defined for Eq. \eqref{eq:dip-mad}. The $d_0$ value is obtained by matching Eq. 
\eqref{eq:ph-p-xs}, using Eq. \eqref{eq:dip-mad}, to the Block et al. \cite{Block:2014fk} structure function, while keeping the remaining parameters as fitted by Block et al. \cite{Block:2014fk} from HERA data.}
\label{tbl:model-param}
\end{table}
\end{center}

\begin{figure}[htbp]
\begin{center}
\hspace{-0.85cm}
\includegraphics[width=0.525\textwidth]{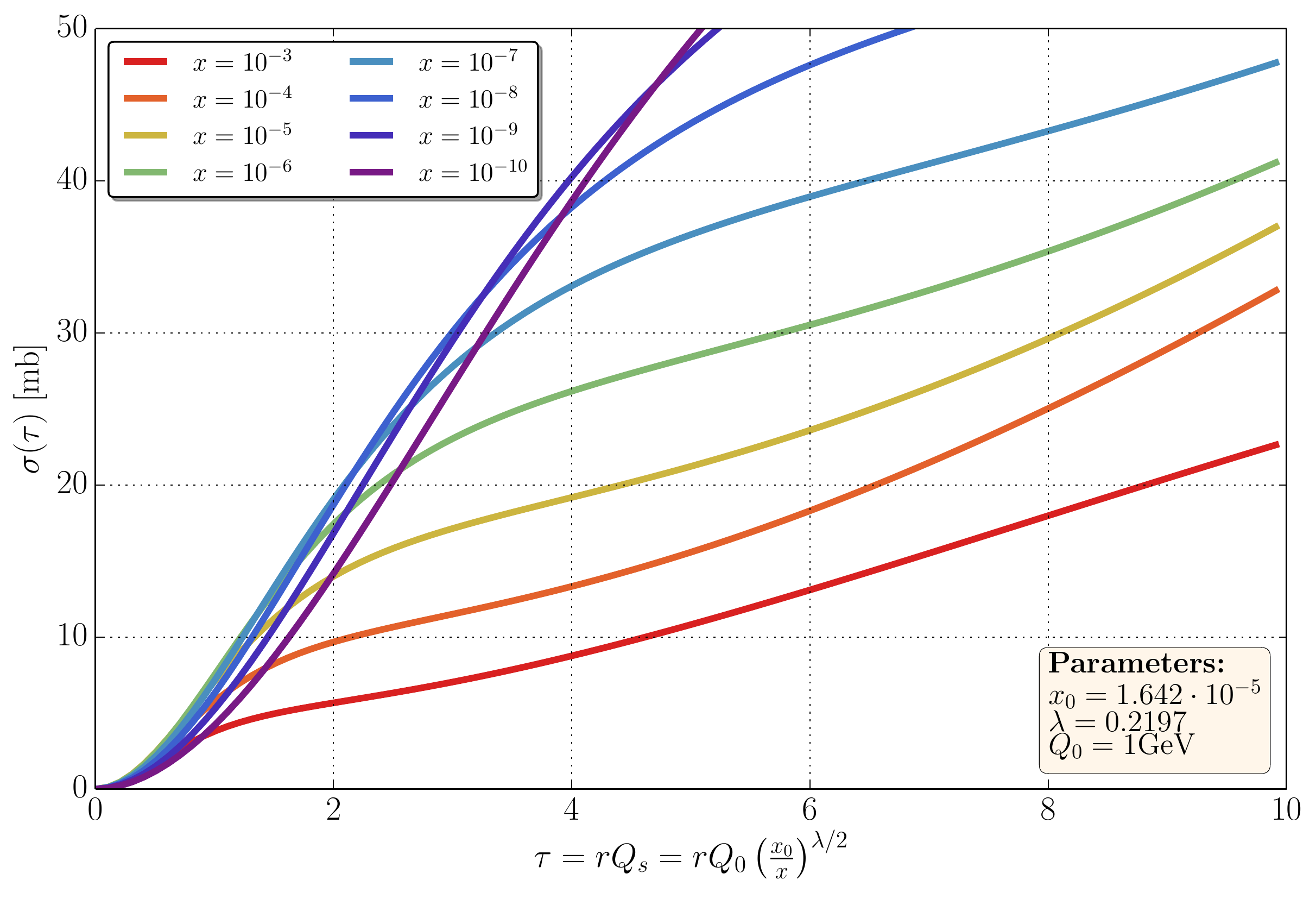}
\caption{Our dipole model for different values of $x$ is shown from $10^{-10}$ (dark purple) to $10^{-3}$ (red) with intermediate lines every power of ten. The dipole cross section monotonically increases as $x$ decreases for large $\tau$.}
\label{fig:dipole-x}
\end{center}
\end{figure}

Our parametrization of ${\sigma}_{dip}$ is determined at low $r$ by the high energy behavior of $F_2$ that correctly incorporates saturation. Also, note that the large-$r$ behavior does not contribute to high energy cross sections because it is suppressed by the wave function in Eq.\ (\ref{eq:ph-p-xs}). In fact, our procedure makes no attempt at describing $F_2$ for values of $x > 10^{-2}$. In Figure \ref{fig:dipole-x} we show the dependence of our dipole parametrization on $x$ and $\tau \equiv r Q_s$, which is related to the color saturation scale $Q_s = Q_0 (x_0/x)^{\lambda/2}$ \cite{Stasto:2000er}.
In order to compare dipole cross sections as a function of $\tau$, we fix the saturation scale parameters to  $Q_0 = 1$ GeV, $x_0 = 1.642 \times 10^{-5}$ and $\lambda = 0.2194$ as in Ref. \cite{Golec-Biernat:zk}. The dipole cross section monotonically increases as $x$ decreases for large $\tau$.

\begin{figure}[htbp]
\begin{center}
\hspace{-0.85cm}
\includegraphics[width=0.525\textwidth]{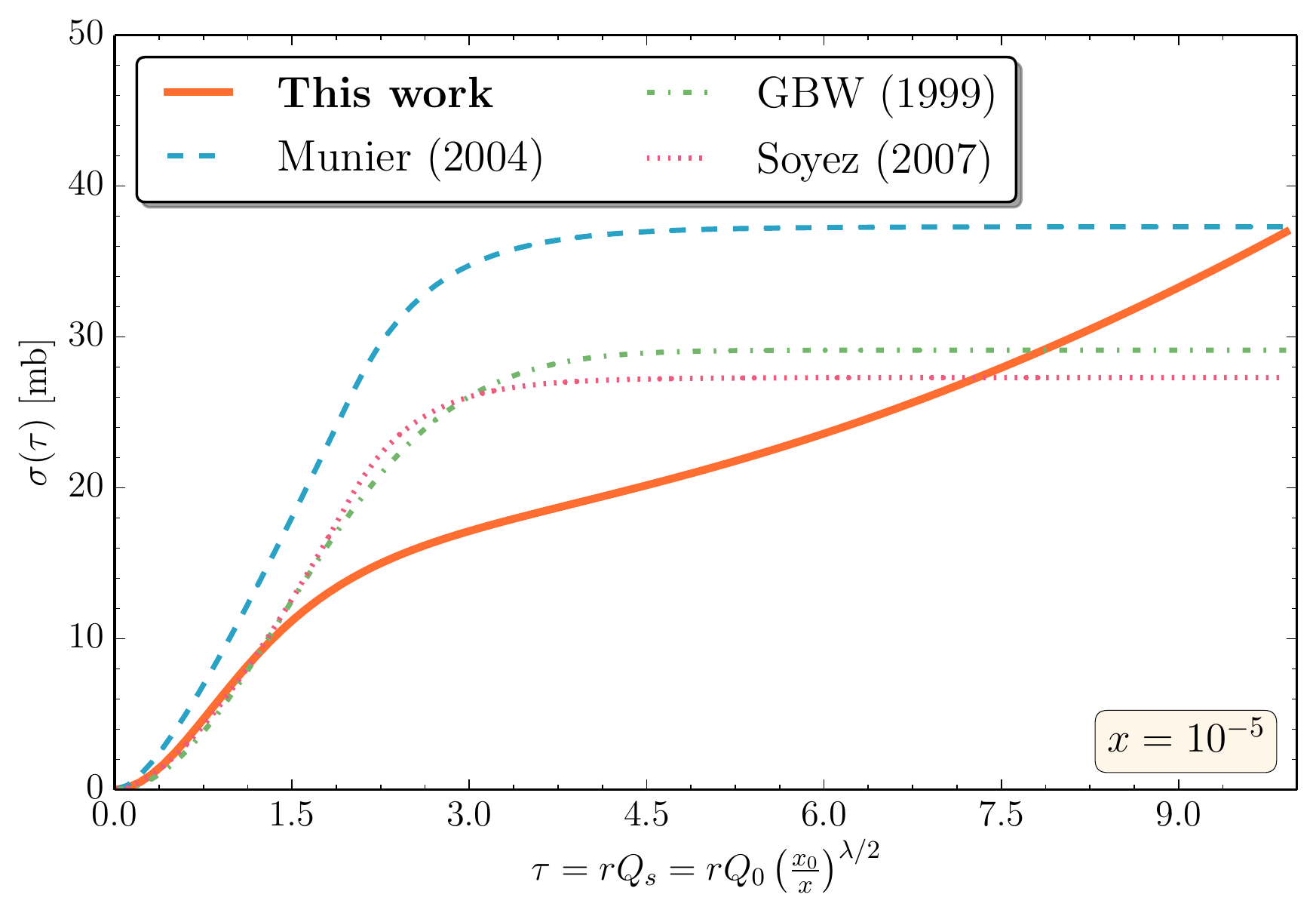}
\caption{Our dipole model compared with other models in the literature  \cite{Golec-Biernat:zk,Munier-note,Soyez:2007kg,Soyez:2006rh} at $x = 10^{-5}$. Note that at small $r$ all the models become color transparent and that at large $r$ they only have linear scale differences.}
\label{fig:dipole-comparison}
\end{center}
\end{figure}

In Figure \ref{fig:dipole-comparison} we compare our dipole cross section for $x=10^{-5}$ with other parameterizations in the literature \cite{Golec-Biernat:zk,Munier-note,Soyez:2007kg}; details can be found in the appendix. The Golec-Biernat-Wusthoff (GBW) dipole \cite{Golec-Biernat:zk} only depends on $\tau$, while the Soyez dipole \cite{Soyez:2007kg} also depends $\ln(1/x)$. Our model for ${\sigma}_{dip}$ displays approximate scaling behavior for large $r$ values where different dipole cross sections vary by less than a factor of two for $x$ in the range of $10^{-3}-10^{-6}$; see also Figure \ref{fig:dipole-x}. 

\begin{figure*}[htbp]
\begin{center}
\includegraphics[width=0.75\textwidth]{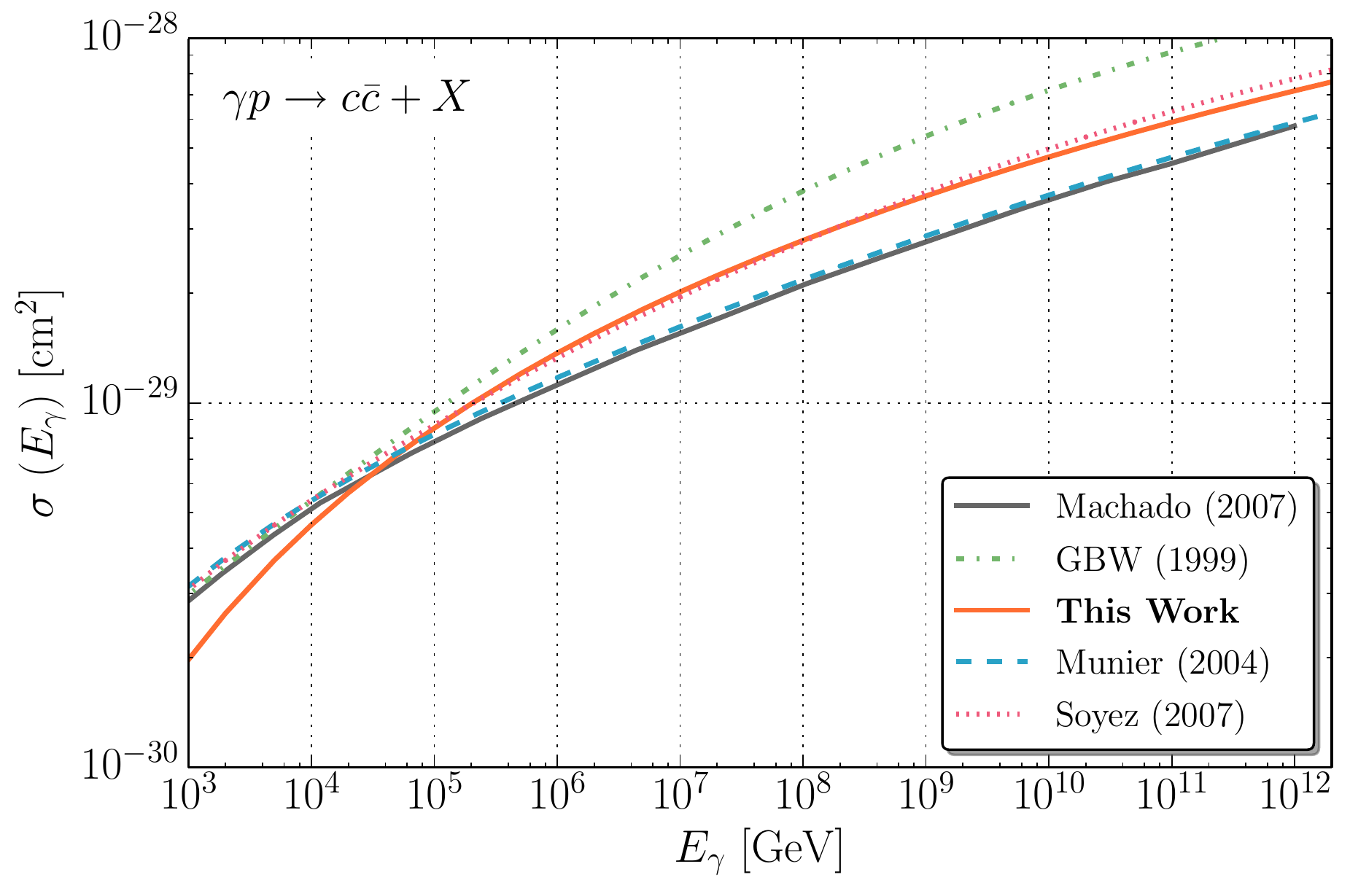}
\caption{Photon-proton charm production as a function of the photon energy is shown for our dipole (upper solid orange) and other dipole models in the literature, namely Munier \cite{Munier-note} (dashed blue), Golec-Biernat et al.\cite{Golec-Biernat:zk} (dot-dashed green), and  Soyez \cite{Soyez:2007kg} (dotted pink) dipole models. A similar calculation by Gonclaves and Machado \cite{Goncalves:2006ch} (lower solid gray) is shown as a reference.}
\label{fig:gamma-proton-xs}
\end{center}
\end{figure*}

At this point we can compute the real photon-proton charm production cross section. In this case $Q^2 \approx 0$, $\sigma^L\approx 0$ and $x \approx (2 m_q)^2/(2m_p E_\gamma)$, the threshold value that dominates the cross section. We fix $m_q=m_c=1.275$ GeV, the Particle Data Group value \cite{PDG:2014}. The result of this calculation is shown in Figure \ref{fig:gamma-proton-xs} where it is compared to the results of other dipole models. The difference between our result (similar to using the Soyez dipole \cite{Soyez:2007kg}) and the Goncalves and Machado \cite{Goncalves:2006ch} result (using the Munier dipole
\cite{Munier-note}) comes from a combination of the choice of charm quark mass and the form of the dipole cross section. The Soyez
and Munier dipoles use the same Iancu-Itakura-Munier form \cite{Iancu:2003ge}, but with different parameters. For example, the values for
$x_0$ differ by two orders of magnitude.

\section{Neutrino-Proton Cross Section}
\label{sec:neutrino-proton-xs} 

\begin{figure}
	\begin{tikzpicture}[line width=1.5 pt, scale=0.75]
		\node at (-0.8, 1.8) {\Large $\nu_\alpha$};
		\draw[fermionout] (-1.0,1.5) -- (0.75,1);
		\node at (3.55, 1.85) {\Large $\nu_\alpha (l_\alpha)$};
		\draw[fermionout] (0.75,1) -- (3.5,1.5);
		\draw[gaugeboson] (0.75,1) -- (2, 0);
		\draw[gluon] (2.8,-0.57) -- (2.8,-1.8);
		
		\draw[fermionin] (3.0,0.6 ) -- (3.5,0.6);
		\draw[connect] (2,0) arc (150:90:1.2);
		
		\draw[fermionout] (3.0,-0.6 ) -- (3.5,-0.6);
		\draw[connect] (2,0) arc (210:270:1.2);
		
		\node at (0.7, 0.2) {\Large $V$};
		\node at (3.2, -1.2) {\Large $g$};
		\node at (3.7, 0.6) {\Large $\bar{q}$};
		\node at (3.7, -0.6) {\Large $q$};
		
		\node at (-1.05, -2.00) {\Large $\mathrm{N}$};
		\node at (4.05, -2.00) {\Large $\mathrm{X}$};
		\draw[fermion] (-0.75,-1.8 ) -- (3.75,-1.8);
		\draw[fermion] (-0.75,-2.0 ) -- (3.75, -2.0);
		\draw[fermion] (-0.75,-2.2 ) -- (3.75,-2.2);
		
		\fill[black] (2,0) circle (.04cm);
		\fill[black] (2.82,-0.575) circle (.04cm);
		
		\fill[kyelloworange]   (2.8,-2)  circle (0.45cm);
		\draw[korange]         (2.8,-2)  circle (0.45cm);
		\node[draw=none,fill=none] at (2.8,-2){$\bf \sigma_{dip}$}; 
		% % % % % % % % % % % % % % % % % % % % % % % %
		
		\node at (5.2, 1.8) {\Large $\nu_\alpha$};
		\draw[fermionout] (5.0,1.5) -- (6.75,1);
		\node at (9.55, 1.85) {\Large $\nu_\alpha (l_\alpha)$};
		\draw[fermionout] (6.75,1) -- (9.5,1.5);
		\draw[gaugeboson] (6.75,1) -- (8, 0);
		\draw[gluon] (8.8,-1.8) -- (8.8,0.6);
		
		\draw[fermionin] (9.0,0.6 ) -- (9.5,0.6);
		\draw[connect] (8,0) arc (150:90:1.2);
		
		\draw[fermionout] (9.0,-0.6 ) -- (9.5,-0.6);
		\draw[connect] (8,0) arc (210:270:1.2);
		
		\node at (6.7, 0.2) {\Large $V$};
		\node at (9.2, -1.2) {\Large $g$};
		\node at (9.7, 0.6) {\Large $\bar{q}$};
		\node at (9.7, -0.6) {\Large $q$};
		
		\node at (4.95, -2.00) {\Large $\mathrm{N}$};
		\node at (10.05, -2.00) {\Large $\mathrm{X}$};
		\draw[fermion] (5.25,-1.8 ) -- (9.75,-1.8);
		\draw[fermion] (5.25,-2.0 ) -- (9.75, -2.0);
		\draw[fermion] (5.25,-2.2 ) -- (9.75,-2.2);
		
		\fill[black] (8,0) circle (.04cm);
		\fill[black] (8.82,-0.575) circle (.04cm);
		
		\fill[kyelloworange]   (8.8,-2)  circle (0.45cm);
		\draw[korange]         (8.8,-2)  circle (0.45cm);
		\node[draw=none,fill=none] at (8.8,-2){$\bf \sigma_{dip}$}; 
	\end{tikzpicture}
\caption{Diagram of $\nu p $ interaction in the dipole picture. In the charge current interaction $V = W^\pm$ whereas in the neutral current $V = Z$. }
\label{fig:diagram-nup}
\end{figure}
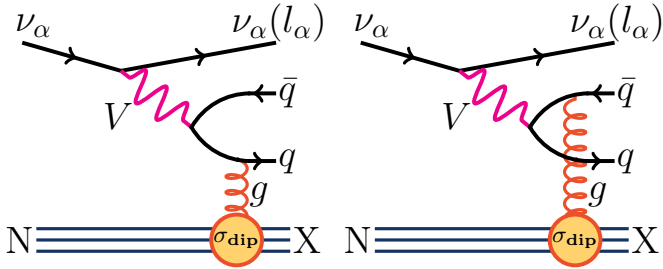

The total neutrino charged (neutral) current cross section for an incident neutrino with energy $E_\nu$ is given by
\begin{equation}
\sigma^{\mathrm{CC}/\mathrm{NC}}_{\nu p}(E_\nu) = \int_{Q^2_{min}}^s dQ^2 \int_{Q^2/s}^1 dx \ \left( \frac{\partial^2 \sigma_{\nu p}}{\partial Q^2 \partial x} \right)_{\mathrm{CC}/\mathrm{NC}}
\label{eq:nu-p-xs}
\end{equation}
with $Q_{min} = 1\  \mathrm{GeV}$ and
\begin{eqnarray}
\frac{\partial^2 \sigma_{\nu N}}{\partial Q^2 \partial x}  &=& \frac{G^2_F}{4\pi x} \left( \frac{M^2_i}{M^2_i + Q^2} \right)^2\nonumber\\
&\times&
\left[ Y_+ F^\nu_T +2 (1-y) F^\nu_L \pm  Y_- x F^\nu_3 \right]
%\left[ Y_+ F^\nu_T(x,Q^2) + (1-y) F^\nu_L(x,Q^2) \pm  Y_- x F^\nu_3(x,Q^2) \right].
\label{eq:dsig2-dq2dx}
\end{eqnarray}
Here $s \approx 2 m_p E_\nu$, $Y_\pm = 1 \pm (1-y)^2$ with $y = Q^2/(xs)$, $M_i = M_W (M_Z)$ for charged (neutral) current interaction, and $``+"$ ($``-"$) refers to neutrinos (antineutrinos). To leading order in pQCD $F^\nu_L = 0$, while $F^\nu_T(x,Q^2)$ and $F^\nu_3(x,Q^2)$ are functions of the parton distribution functions, e.g., for a neutrino charged (CC) and neutral current (NC) interaction with an isoscalar target nucleon $N$ \cite{Cooper-Sarkar:2011fc}
\begin{equation}
\mathrm{CC:} \left\{
\begin{aligned}
F^\nu_T = x (u + \bar{u} + d +\bar{d}  + 2s + 2b + 2 \bar{c}) \\
x F^\nu_3 = x (u - \bar{u} + d -\bar{d} + 2s + 2b  -  2 \bar{c}) 
\end{aligned}
\right. ~,
\label{eq:F2neu-CC}
\end{equation}
%
%\begin{equation}
%\mathrm{NC:} \left\{
%\begin{aligned}
%F^\nu_T &=& x \left[ (1/2 - s^2_w + 10/9\ s^4_w)(u  + \bar{u}+ d  +\bar{d}) \right. \\
%		& &+ (1/2 - 4/3\ s^2_w + 16/9\ s^4_w)(c  + \bar{c}) \\
%		& & \left. + (1/2 - 2/3\ s^2_w + 4/9\ s^4_w)(s + b  + \bar{s} +  \bar{b})  \right] \\
%x F^\nu_3 &=& x \left[ (1/2 - s^2_w) (u - \bar{u} + d - \bar{d}) \right]
%\end{aligned}
%\right. ~,
%\label{eq:F2neu-NC}
%\end{equation}
%
\begin{equation}
\mathrm{NC:} \left\{
\begin{aligned}
F^\nu_T &= x \left[ \frac{1}{4}(L_u^2+R_u^2+L_d^2+R_d^2)) (u  + \bar{u}+ d  +\bar{d}) \right. \\
		& + \frac{1}{2}(L_u^2+R_u^2)(c  + \bar{c}) \\
		&  \left. + \frac{1}{2}(L_d^2+R_d^2)(s + b  + \bar{s} +  \bar{b})  \right] \\
x F^\nu_3 &= x \left[ \frac{1}{2}(L_u^2-R_u^2+L_d^2-R_d^2) (u - \bar{u} + d - \bar{d}) \right]
%compare this F3 to above to see factor of 2 difference
\end{aligned}
\right. % ~,
\label{eq:F2neu-NC}
\end{equation}
where $s_w = \sin(\theta_w)$ is the sine of the weak mixing angle and the weak couplings are given by
$L_u=1-4/3 s_w^2,\ L_d=-1+2/3 s_w^2,\ R_u=-4/3 s_w^2$ and $R_d=2/3 s_w^2$.
 The antineutrino structure functions for charged and neutral current are obtained by replacing $q \to \bar{q}$ and $F^{\bar{\nu}}_3 \to - F^{\nu}_3 $. For high neutrino energy, i.e. small $x$, the $F_3$ term does not contribute.

We evaluate $F^\nu_{L/T}$ using the dipole formalism with
\begin{equation}
F^\nu_{T/L} = \frac{Q^2}{4 \pi^2} \sum_q \int_0^1 dz \int d^2{\bf r} | \psi^{W,Z}_{T/L,q} (z,{\bf r};Q^2)|^2 \sigma_{dip} (x,{\bf r})
 ~,
\label{eq:FTL-dipole}
\end{equation}
where $\psi^{W,Z}_{T/L,q}(z,{\bf r};Q^2)$ corresponds to the wave function for a vector boson ($W$ or $Z$), of virtual momenta $Q^2$,  to fluctuate into a $q\bar{q}$ pair with fractional longitudinal momentum $z$ and transverse spatial separation $r$. They are computed from the diagrams in Figure \ref{fig:diagram-nup}; in the massless quark limit \cite{Barone1993176,Kutak:2003bd}
% HR: changed these equations, originally written this way in earlier version
%\begin{eqnarray}
%| \psi^{W,Z}_{T,q} (z,{\bf r};Q^2)|^2 &=& \frac{N_c}{2 \pi^2} Q^2 \left[z \bar{z} \right] \left[ z^2 + \bar{z}^2 \right] K_1(\epsilon r) ,~\\ 
%| \psi^{W,Z}_{L,q} (z,{\bf r};Q^2)|^2 &=& \frac{2 N_c}{\pi^2} Q^2 \left[ z \bar{z} \right] K_0(\epsilon r) 
%~.
%\label{eq:psi-TL}
%\end{eqnarray}
\begin{eqnarray}
| \psi^{W}_{T,q} (z,{\bf r};Q^2)|^2 &=& \frac{2 N_c}{ \pi^2} Q^2 \left[z \bar{z} \right] \left[ z^2 + \bar{z}^2 \right] K_1^2(\epsilon r) ,~\\ 
| \psi^{W}_{L,q} (z,{\bf r};Q^2)|^2 &=& \frac{8 N_c}{\pi^2} Q^2 \left[ z \bar{z} \right] ^2K^2_0(\epsilon r) \ ,\\
\nonumber
| \psi^{Z}_{T,q} (z,{\bf r};Q^2)|^2 &=& \frac{ N_c}{ 2\pi^2} [L_u^2+L_d^2+R_u^2+R_d^2]\\
&\times & Q^2 \left[z \bar{z} \right] \left[ z^2 + \bar{z}^2 \right] K_1^2(\epsilon r) ,~\\ 
\nonumber
| \psi^{Z}_{L,q} (z,{\bf r};Q^2)|^2 &=& \frac{2 N_c}{\pi^2} [L_u^2+L_d^2+R_u^2+R_d^2]\\
&\times &
Q^2 \left[ z \bar{z} \right] ^2K^2_0(\epsilon r) 
~.
\label{eq:psi-TL}
\end{eqnarray}
Following the Henley et al. \cite{Henley:2006lp} prescription, we switch from the pQCD parametrization of the structure functions and the high energy dipole formalism by using Eqns. \eqref{eq:F2neu-CC} and \eqref{eq:F2neu-NC} for $x < x_0$ and Eq. \eqref{eq:FTL-dipole} otherwise.

\begin{figure*}[htbp]
\begin{center}
\includegraphics[width=0.75\textwidth]{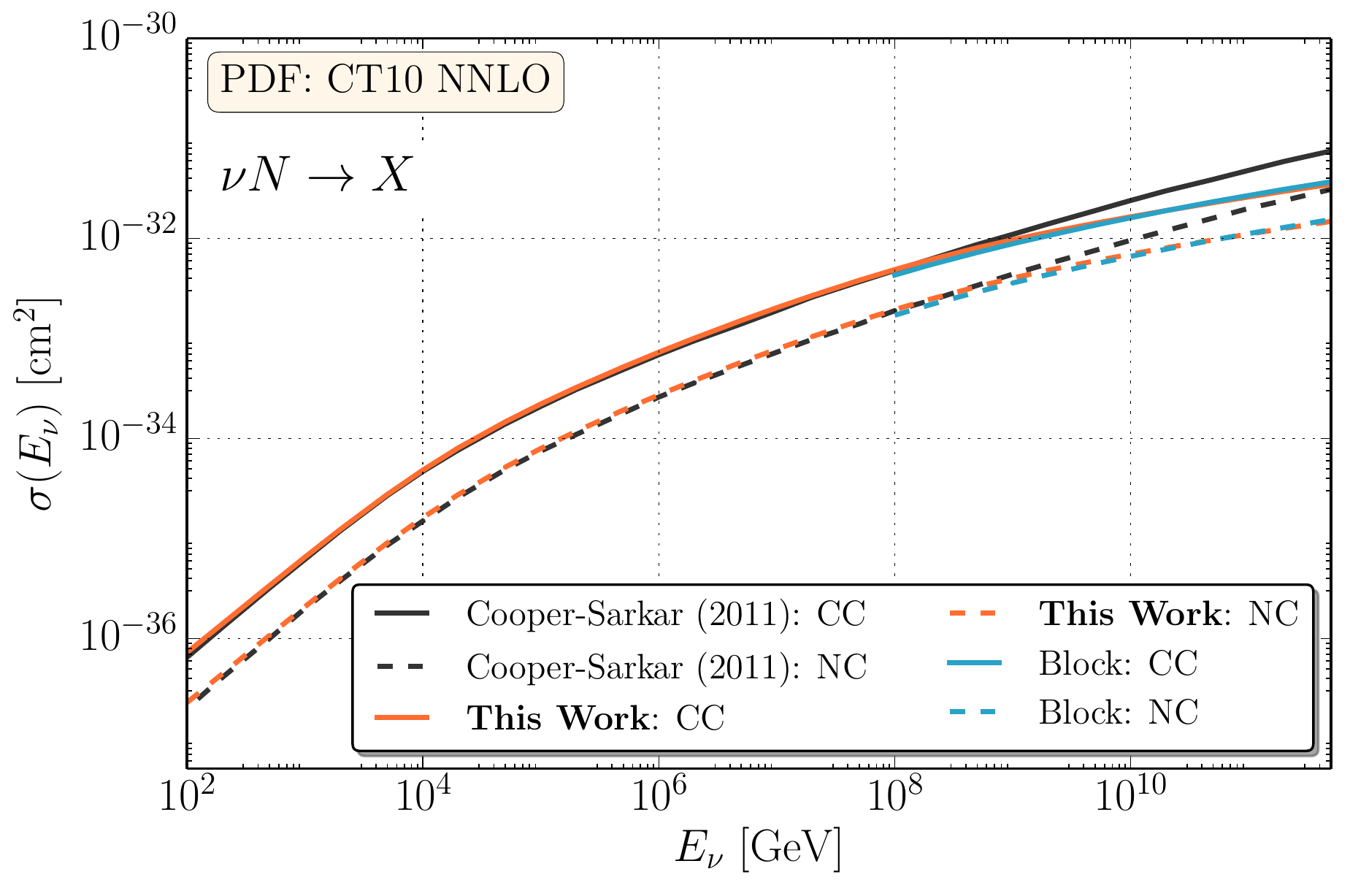}
\caption{Neutrino cross sections calculated using HERAPDF NNLO PDFs \cite{HERA:2009} and our dipole model (orange). For comparison, we show a recent calculation done in pQCD by Cooper-Sakar et al. \cite{Cooper-Sarkar:2011fc} (black), which does not incorporate saturation effects and a similar calculation by Block et al. \cite{Block:2014fk}. The solid lines represent the charge current cross sections, whereas the dashed line portray the neutral current cross section.}
\label{fig:neutrino-proton-xs}
\end{center}
\end{figure*}

Our result for the total charged current cross section is shown in Figure \ref{fig:neutrino-proton-xs}. We used the CT10nnLO PDFs \cite{Gao:2013xoa} and $x_0 = 10^{-5}$. At low energies our calculation agrees with the pQCD calculation, but it incorporates the saturation effect at ultrahigh energies that reduce the neutrino cross section for $E_\nu > 10^9 \ {\rm GeV}$.
The ultrahigh energy cross section is not very sensitive to the choice of $x_0$ between $10^{-2}-10^{-6}$. Our calculation 
with $x_0=10^{-5}$ can be directly compared to the results of Block et al. \cite{Block:2014fk}. As a reference, we show the Cooper-Sakar et al. \cite{Cooper-Sarkar:2011fc} calculation without saturation effects; see also \cite{Jeong:2010za,JimenezDelgado:2009tv,Connolly:2011vc}.
In Table \ref{tbl:nuCCTbl}, we have tabulated the charge current cross section for the three calculations shown in Figure \ref{fig:neutrino-proton-xs}.
 %changed  Enberg ref to Soyez
 
\begin{table}[htbp]
\centering
\begin{tabular}{ | l | c | c | c | }
\hline
\hline
  \multirow{2}{*}{$E_{\nu} [ \text{GeV} ]$} & \multicolumn{3}{c|}{$\sigma_{\text{CC}} [ \text{cm}^2 ]$} \\ \cline{2-4}
  & Sarkar et al. & Block et al. & \bf{This work} \\
\hline
  $1 \times 10^8$ & $4.8 \times 10^{-33}$ & $4.31 \times 10^{-33}$  & $4.88 \times 10^{-33}$ \\
  $2 \times 10^8$ & $6.2 \times 10^{-33}$ & $5.46 \times 10^{-33}$ & $6.15 \times 10^{-33} $\\
  $5 \times 10^8$ & $8.7 \times 10^{-33}$ & $7.25 \times 10^{-33}$ & $8.07 \times 10^{-33} $\\
  $1 \times 10^9$ & $1.1 \times 10^{-32}$ & $8.87 \times 10^{-33}$ & $9.71 \times 10^{-33} $\\
  $2 \times 10^9$ & $1.4 \times 10^{-32}$ & $1.07 \times 10^{-32}$ & $1.15 \times 10^{-32} $\\
  $5 \times 10^9$ & $1.9 \times 10^{-32}$ & $1.36 \times 10^{-32}$ & $1.42 \times 10^{-32} $\\
  $1 \times 10^{10}$ & $2.4 \times 10^{-32}$ & $1.61 \times 10^{-32}$ & $1.65 \times 10^{-32} $\\
  $2 \times 10^{10}$ & $3.0 \times 10^{-32}$ & $1.90 \times 10^{-32}$ & $1.90 \times 10^{-32} $\\
  $5 \times 10^{10}$ & $3.9 \times 10^{-32}$ & $2.33 \times 10^{-32}$ & $2.28 \times 10^{-32} $\\
  $1 \times 10^{11}$ & $4.8 \times 10^{-32}$ & $2.69 \times 10^{-32}$ & $2.60 \times 10^{-32} $\\
  $2 \times 10^{11}$ & $5.9 \times 10^{-32}$ & $3.10 \times10^{-32}$ & $2.96 \times 10^{-32} $\\
  $5 \times 10^{11}$ & $7.5 \times 10^{-32}$ & $3.69\times10^{-32}$ & $3.49 \times 10^{-32} $\\
\hline
\end{tabular}
\caption{Neutrino charge current cross section values for Sarkar et. al \cite{Cooper-Sarkar:2011fc} pQCD calculation, Block et al. \cite{Block:2014fk}, and this work.}
\label{tbl:nuCCTbl}
\end{table}

\section{Proton-Proton Cross Section}
\label{sec:proton-proton-xs} 

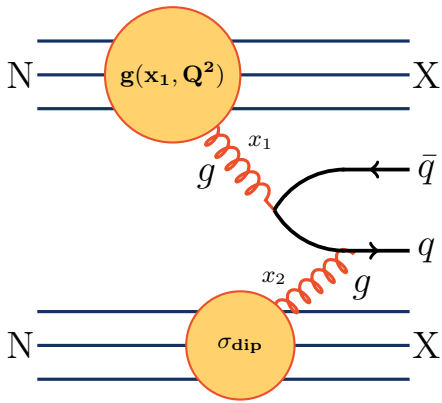
\begin{figure}
\begin{tikzpicture}[scale=0.90,
					thick,
			% Set the overall layout of the tree
			level/.style={level distance=3.15cm, line width=0.8mm},
			level 2/.style={sibling angle=60},
			level 3/.style={sibling angle=60},
			level 4/.style={level distance=1.4cm, sibling angle=60}
	]

	% lower proton
	\node at (-4.25, -2.) {\Large $\mathrm{N}$};
	\node at (1.75, -2.00) {\Large $\mathrm{X}$};
	\draw[fermion] (-4,-2.5) -- (1.5,-2.5);
	\draw[fermion] (-4,-1.5) -- (1.5,-1.5);
	\draw[fermion] (-4,-2)    -- (1.5,-2);
	% upper proton
	\node at (-4.25, 2.) {\Large $\mathrm{N}$};
	\node at (1.75, 2.00) {\Large $\mathrm{X}$};
	\draw[fermion] (-4,2)    -- (-2,2);
	\draw[fermion] (-4,2.5) -- (-2,2.5);
	\draw[fermion] (-4,1.5) -- (-2,1.5);
	
	% gluons
	\node[draw=none,fill=none] at (-0.7,1){$x_1$}  ; 
	\node at (-1.5, 0.5) {\Large $g$};
	\draw[gluon]   (-2,1.8) --     (-0.5,0) ;
	\node[draw=none,fill=none] at (-0.5,-1){$x_2$}  ; 
	\node at (0.8,-1.15) {\Large $g$};
	\draw[gluon]      (-1,-2) --    (0.7,-0.6) ;
	
	% dipole
	\draw[fermionin, line width = 0.5 mm] (0.5,0.6 ) -- (1.5,0.6);
	\draw[connect, line width = 0.5 mm] (-0.5,0) arc (150:90:1.2);
	
	\draw[fermionout, line width = 0.5 mm] (0.5,-0.6 ) -- (1.5,-0.6);
	\draw[connect, line width = 0.5 mm] (-0.5,0) arc (210:270:1.2);
	
	\node at (1.75, 0.6) {\Large $\bar{q}$};
	\node at (1.75, -0.6) {\Large $q$};
	
	%Other quarks
	\draw[fermion]   (-2,2)  --  (1.5,2) ;
	\draw[fermion]   (-2,2.5) --  (1.5,2.5) ;
	\draw[fermion]   (-2,1.5) --  (1.5,1.5) ;
	%\draw[fermion]   (-2,-2) --  (2,-3) ;
	%\draw[fermion]   (-2,-2.5) -- (2,-3.5) ;
	
	%pdf circle
	\fill[kyelloworange]   (-2,2)  circle (1cm);
	\draw[korange]         (-2,2)  circle (1cm);
	\node[draw=none,fill=none] at (-2,2){$\bf g(x_1,Q^2)$}; 
	
	\fill[kyelloworange]   (-1,-2)  circle (0.8cm);
	\draw[korange]         (-1,-2)  circle (0.8cm);
	\node[draw=none,fill=none] at (-1,-2){$\bf \sigma_{dip}$}; 

\end{tikzpicture}
\caption{Diagram of $p p \to q \bar{q} X $ for one dipole interaction topology, the remainder are shown in Figure \ref{fig:diagram-dipole-gluon} . In this picture the gluon from the projectile proton has fractional parton momenta $x_1$ and it is modeled through PDF, while the target proton gluon has fractional parton momenta $x_2 \ll 1$ and it is modeled through a dipole interaction.}
\label{fig:diagram-proton}
\end{figure}

\begin{figure}
	\begin{tikzpicture}[line width=1.3 pt, scale=0.75]
		\draw[gluon] (0,0) -- (1, 0) ;
		\draw[gluon] (1,0) -- (2, 0) ;
		\draw[gluon] (1,0) -- (1, -1.6) ;

		\draw[fermionin] (3.0,0.6 ) -- (3.5,0.6);
		\draw[connect] (2,0) arc (150:90:1.2);
		
		\draw[fermionout] (3.0,-0.6 ) -- (3.5,-0.6);
		\draw[connect] (2,0) arc (210:270:1.2);

		\node at (1, 0.5) {\Large $g$};
		
		\node at (3.7, 0.6) {\Large $\bar{q}$};
		\node at (3.7, -0.6) {\Large $q$};
		
		\fill[black] (2,0) circle (.05cm);
		
		\fill[black] (1,0) circle (.05cm);
		% % % % % % % % % % % % % % % % % % % % % % % %
		\draw[gluon] (4.5,0) -- (6, 0);
		\draw[gluon] 	  (6.8,0.57) -- (6.8,-1.6);
		
		\draw[fermionin]  (7.0,0.6 ) -- (7.5,0.6);
		\draw[connect]  (6,0) arc (150:90:1.2);
		
		\draw[fermionout] (7.0,-0.6 ) -- (7.5,-0.6);
		\draw[connect] (6,0) arc (210:270:1.2);
		
		\node at          (5, 0.5) {\Large $g$};
		\node at 		  (7.25, -0) {\Large $g$};
		\node at 		  (7.7, 0.6) {\Large $\bar{q}$};
		\node at 		  (7.7, -0.6) {\Large $q$};
		
		\fill[black]      (6,0) circle (.05cm);
		\fill[black]      (6.82,0.575) circle (.05cm);
				
		% % % % % % % % % % % % % % % % % % % % % % % % % % % %
		
		\draw[gluon] (8.5,0) -- (10, 0);
		\draw[fermionin] (11,0.6 ) -- (11.5,0.6);
		\draw[connect] (10,0) arc (150:90:1.2);
		
		\draw[fermionout] (11, -0.6) -- (11.5,-0.6);
		\draw[connect] 	  (10, 0) arc(210:270:1.2);
		
		\draw[gluon] 	(10.8, -0.6) -- (10.8, -1.6);
		
		\node at (11.25, -1.2) {\Large $g$};
		
		\fill[black] (10.82,-0.575) circle (.05cm);
		\fill[black] (10,-0) circle (.05cm);
		\node at (11.7, 0.6) {\Large $\bar{q}$};
		\node at (11.7, -0.6) {\Large $q$};
	\end{tikzpicture}
\caption{Possible $g \to q \bar{q}$ interactions with a gluon in the dipole picture.}
\label{fig:diagram-dipole-gluon}
\end{figure}
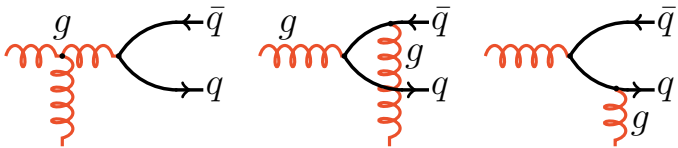

In the dipole model the proton-proton cross section is given by 
\cite{Nikolaev:1994de,Nikolaev:1993th}
\begin{eqnarray}
\hspace{-0.25cm}
\nonumber
\sigma_{pp \to q\bar{q} +X }&= & 2 \int^{-\ln(2m_q/\sqrt{s})}_0 dy \ x_1 g(x_1, \mu)\\
&\times & {\sigma}_{gN \to q\bar{q} + X}(x_2;Q^2)
 ~,
\label{eq:pp-1}
\end{eqnarray}
where $g(x_1,\mu)$ is the gluon distribution function at the scale $\mu$,  and $x_{1,2}$ satisfy $x_1 x_2 \simeq (2 m_q)^2/s$
with
\begin{eqnarray*}
x_1&\simeq & \frac{2m_q}{\sqrt{s}}\exp(y)\\
x_2&\simeq & \frac{2m_q}{\sqrt{s}} \exp(-y);
\end{eqnarray*}
see Figure \ref{fig:diagram-proton}. Here ${\sigma}_{gN \to q\bar{q} + X}$ is the partonic cross section, which in the dipole picture is given by
\begin{eqnarray}
{\sigma}_{gN \to q\bar{q} + X}(x_2;Q^2) &=&  \int dz\ d^2 {\bf r}\  | \psi^{g}_{T,q} (z,{\bf r};Q^2)|^2 \nonumber
\\ &\times & {\sigma}_{gq\bar{q}} (x_2,z,{\bf r})
\label{eq:pp-2}
\end{eqnarray}
%HR: slight reformatting of above equation
The partonic cross section is directly related to the dipole cross section \cite{Nikolaev:1994de,Nikolaev:1993th}:
%JD: it was "partonic cross sector" and I assumed it should be "partonic cross section"
%
\begin{eqnarray}
\hat{\sigma}_{gq\bar{q}} (x_2,z,{\bf r})&=& \frac{9}{8} \left[ \sigma_{dip}(x_2,z {\bf r}) + \sigma_{dip}(x_2,\bar{z} {\bf r}) \right] 
\nonumber \\
&-& \frac{1}{8} \sigma_{dip}(x_2, {\bf r})
\label{eq:gqq}
\end{eqnarray}
where the different terms correspond to the superposition of the diagrams shown in Figure \ref{fig:diagram-dipole-gluon}. The gluon wave function is related to the photon wave function, given in Eq. \eqref{eq:ph-wavefunction-T} by \cite{Nikolaev:1995ty}
\begin{equation}
| \psi^{g}_{T,q} (z,{\bf r};Q^2)|^2 = \frac{\alpha_s}{N_c \alpha_{em}}| \psi^{\gamma}_{T,q} (z,{\bf r};Q^2)|^2.
\end{equation}

\begin{figure*}
\begin{center}
\includegraphics[width=0.6\textwidth]{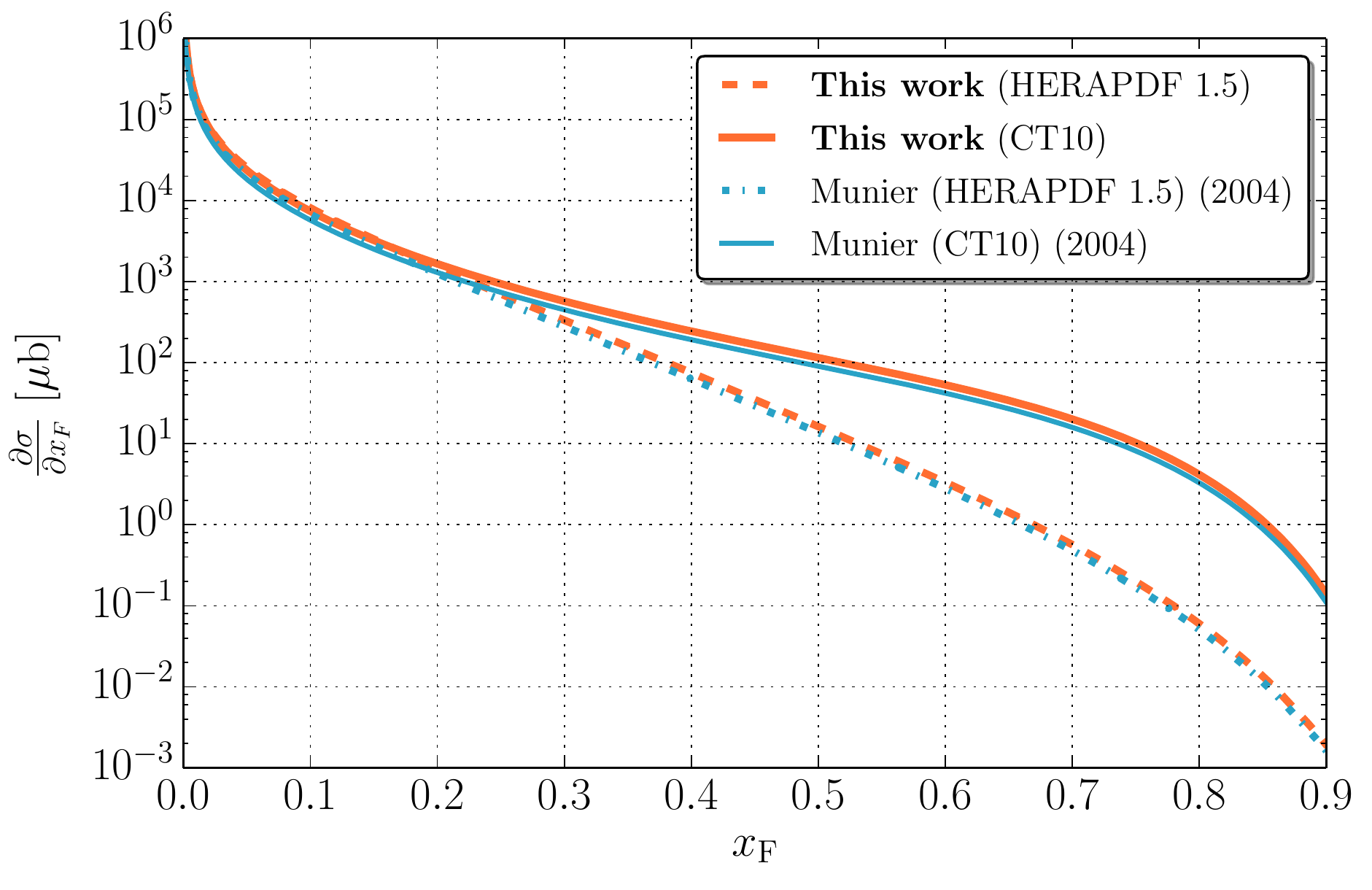}
\caption{Proton-proton charm production differential cross section as a function of Feynman $x_F$ at incident proton energy of $10^9 \mathrm{GeV}$ is shown for our (orange) and Munier \cite{Munier-note} (blue) dipoles. Furthermore, solid lines are calculations done using HERAPDF 1.5 \cite{HERA:2009} while dashed lines uses CT10 NNLO\cite{Gao:2013xoa}.}
\label{fig:proton-proton-to-cc-xf}
\end{center}
\end{figure*}

\begin{figure*}
\begin{center}
\includegraphics[width=0.6\textwidth]{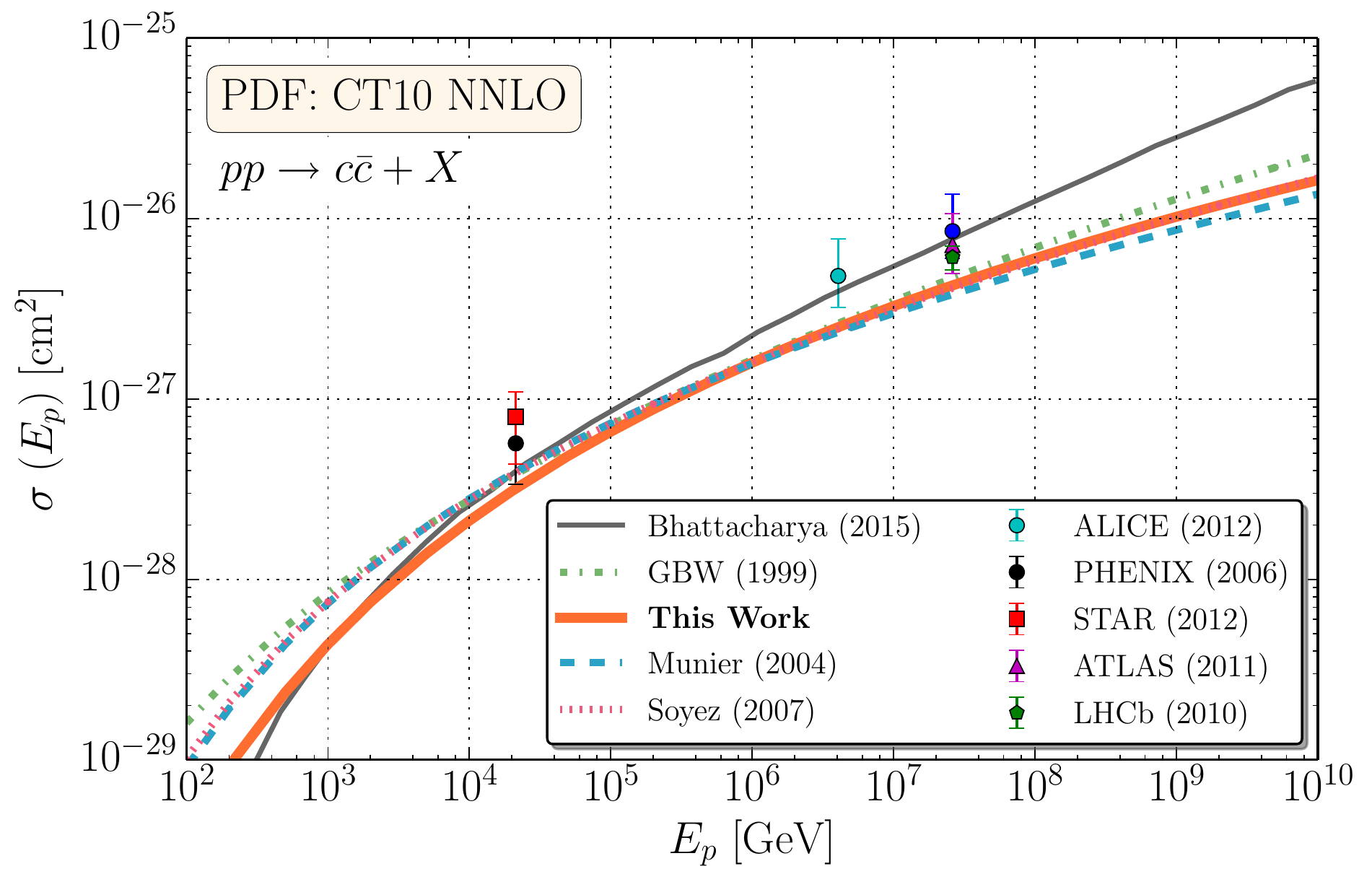}
\caption{Proton-proton charm production deep inelastic cross section as a function of the proton energy is shown for our dipole (orange), and other dipole models in the literature, namely Golec-Biernat and Wusthoff \cite{Golec-Biernat:zk} (green), Munier (blue) \cite{Munier-note} and Soyez \cite{Soyez:2007kg} (pink). In all cases, we have used CT10 NNLO PDFs \cite{Gao:2013xoa} parton distribution functions. Furthermore, a recent calculation by Bhattacharya et al. \cite{Bhattacharya:2015jpa} (gray) is shown as a reference as well as recent experimental results from the ALICE collaboration \cite{Abelev:2012vra}.}
\label{fig:proton-proton-to-cc}
\end{center}
\end{figure*}

We first calculate the cross section for producing charm particles. While such a calculation is not new, our dipole calculational framework is directly constrained by a wealth of data, and can even accomodate the total cross section as we will see next. Atmospheric neutrino measurements mostly sample the very forward, large Feynman-$x_F$, cross section; we therefore show in Figure \ref{fig:proton-proton-to-cc-xf} our results for the $d\sigma (p p \to c \bar{c})/dx_F$ distribution. We find that differences resulting from different dipole parametrizations are smaller than the spread associated with different PDFs. In Figure \ref{fig:proton-proton-to-cc} we have plotted the total proton-proton charm production cross section for different dipole models using the CT10nnLO gluon PDF with $\mu = \sqrt{4 m^2_c}$ and $\alpha_s = 0.33$. The high energy
dipole model evaluations of
$\sigma  (p p \to c \bar{c} X)$ are lower than the central perturbative evaluation of the charm pair
cross section shown in, e.g. \cite{Bhattacharya:2015jpa}. The perturbative calculation has large uncertainties associated with the scale
dependence and for $E_p=10^{10}$ GeV and scales dependent on factors of $m_c$, the uncertainty
band drops to $\sigma\sim 10^{-26}$ cm$^2$, the level of the dipole prediction.
%the highest energy shown in Fig. \ref{fig:proton-proton-to-cc}, 

\begin{figure*}[htbp]
\begin{center}
\includegraphics[width=0.75\textwidth]{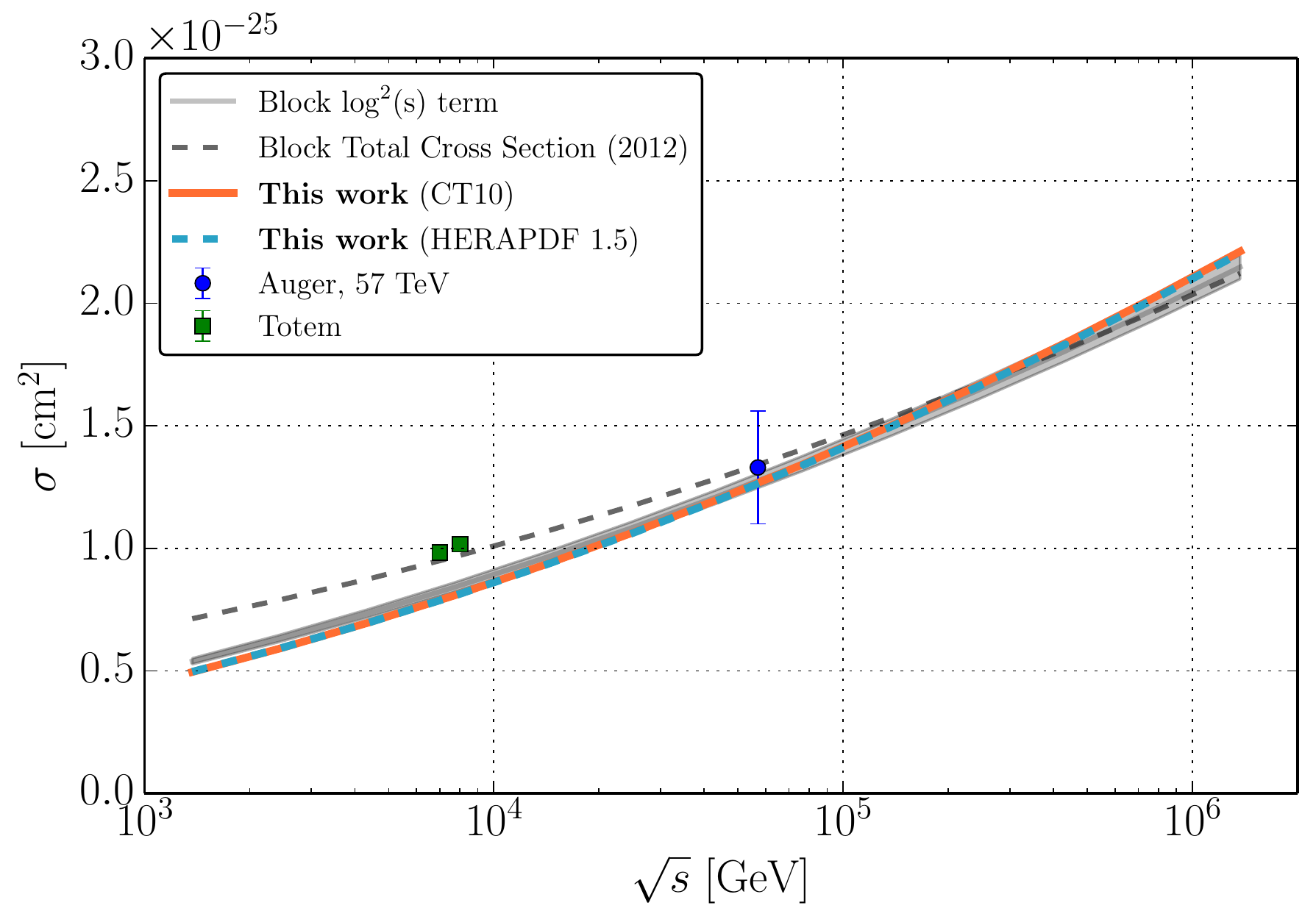}
\caption{Best fit total proton-proton cross section calculated using the dipole approximation with CT10NLO \cite{Nadolsky:2008zw} (solid orange) and HERAPDF1.5NLO \cite{HERA:2009} (dashed light blue) PDFs compared to Block et al. \cite{Block:2011ak} (dashed gray). Furthermore, we include only the $log^2(s)$ term from Block et al. fit (solid gray line), which dominates the high energy cross section and is in agreement with our calculation; the associated gray band is the one sigma error given by the Block et al. fit.}
\label{fig:total-ppxs}
\end{center}
\end{figure*}

Finally, we can use Eqns. (\ref{eq:pp-1}-\ref{eq:gqq}) to calculate the {\it asymptotic} proton-proton total cross section. We substitute \cite{Nikolaev:1990ja} Eq. \eqref{eq:gqq} by \cite{Nikolaev:2005zj}
\begin{equation}
\hat{\sigma}_{ggg} = \frac{1}{2} \left[ \sigma_{dip}(x_2,z {\bf r}) + \sigma_{dip}(x_2,\bar{z} {\bf r})  +  \sigma_{dip}(x_2, {\bf r}) \right] \\
\label{eq:ggg-1}
\end{equation}
\begin{equation}
| \psi^{g}_{T,gg} (z,{\bf r};Q^2)|^2 = 2 ( N_c -1 ) | \psi^{g}_{T,q} (z,{\bf r};Q^2)|^2 ~,
\label{eq:ggg-2}
\end{equation}
and replace $m_q \to m_g$, an {effective gluon mass}, in evaluating $x_1$ and $x_2$. The dipole picture correctly reproduces the $\ln^2(s)$-dependence of the $pp$ total cross section at high energies. We match the prediction of Ref. \cite{Block:2011ak} that at ultra-high energies the total cross section 
\begin{equation}
\sigma_{pp}(s) \sim (0.2817) \ln^2(s/(2 m^2_p)) ~ \mathrm{mb}
 ~,
\label{eq:bloch-halzen}
\end{equation}
in agreement with the LHC \cite {Antchev:2013gaa} 
and Auger \cite{Collaboration:2012wt} measurements; see Figure \ref{fig:total-ppxs}. We fit for $m_g$ and $\mu$, while using $\alpha_s = \alpha_s(\mu)$, and yield $ m_g = 0.154 ~ \mathrm{GeV}$ with $\mu^2  = 1.69 ~ {\rm GeV}^2$ ($\mu^2  = 1.6 ~ {\rm GeV}^2$) for CT10 NLO\cite{Nadolsky:2008zw} (HERAPDF1.5NLO \cite{HERA:2009}). We plot the resulting fits in Figure \ref{fig:total-ppxs}, where we also shown, for comparison, Block et al. \cite{Block:2011ak} calculation; at the highest energies where the cross section is dominated by the $\ln^2(s)$ term good agreement is found between our calculation and Block et al.

Because the asymptotic cross sections are dominated by gluons only, our formalism predicts that all cross sections become equal with increasing energy, independent of the quantum numbers of the projectile and target.

\section{Conclusion}
\label{sec:conclu} 

We have used a theoretically motivated but approximate form of the dipole cross section that extends perturbative QCD calculations into the small-$x$ regime. Ambiguities associated with the small-$x$ behavior of the structure functions are mitigated by requiring a saturation that reproduces cross sections that behave asymptotically as  $\ln^2(s)$, in agreement with data.
Using Eq. \eqref{eq:dip-mad} to relate $F_2(x,Q^2)$ to $\sigma_{dip}(x,r)$, we have used the structure function parameterization of
Ref. \cite{Block:2014fk} that reflects this asymptotic behavior. Using this formalism, we have calculated the high-energy $\gamma p$, $\nu p$ and $pp$ cross sections. Our results agree remarkably well with other dipole model calculations, and we predict the very high-energy behavior of the neutrino cross section in the EeV energy range as well as the hadronic production of charm pairs and total $pp$ cross section. 
These are of interest to neutrino telescopes and high energy cosmic ray experiments.

\section*{Acknowledgments}

The authors acknowledge support from the Wisconsin IceCube Particle Astrophysics Center (WIPAC). C.A., F.H., and L.W. were supported in part by the
U.S. National Science Foundation under Grants No. OPP-0236449 and PHY-
0969061 and by the University of Wisconsin Research Committee with funds
granted by the Wisconsin Alumni Research Foundation.  
M.K. was funded by the Martin-Schmeier-Stiftung scholarship. M.H.R. is supported in part by the U.S. Department of Energy 
grant DE-SC0010114.

\section{Appendix}
\label{sec:appendix} 

Here, we specify the dipole models used for comparison in the text. In Figure \ref{fig:dipole-comparison}, we plot the different dipoles together assuming the values given in their respective references.

\subsection{The GBW dipole}

This dipole cross section defined by Golec-Biernat and Wusthoff in Ref. \cite{Golec-Biernat:zk} is
\begin{equation*}
\sigma^{\rm GBW}_{dip}(x,r) = \sigma_0 \left( 1 - \exp \left[ - \frac{1}{4} r^2 Q^2_s(x) \right] \right)
 ~,
\label{eq:dip-gbw}
\end{equation*}
where the  saturation scale is given by $Q_s(x) = Q_0 \left( \frac{x_0}{x}\right)^{\lambda /2}$ and the parameter values have been fixed at $Q_0=1$ GeV, $\lambda = 0.277$, $x_0 = 4.1\times 10^{-5}$ and $\sigma_0 = 29.12\ {\rm mb}$.

\subsection{The Munier and Soyez dipoles}

This dipole cross sections of Ref. \cite{Munier-note} and
Soyez \cite{Soyez:2007kg} is of the Iancu-Itakura-Munier form \cite{Iancu:2003ge},
\begin{equation*}
\sigma^{\rm Munier/Soyez}_{dip}(x,r) = \sigma_0 N(\tau,Y)
 ~,
\label{eq:dip-gbw}
\end{equation*}
where 
\begin{equation*}
N(\tau,Y) =  \left\{
\begin{aligned}
N_0 (\tau/2)^{2 \gamma_{eff}(x,r)} , \tau < 2\\
1 - \exp[-a \ln^2(b \tau) ] , \tau \geq 2
\end{aligned}
\right. ~,
\label{eq:dip-gbw-N}
\end{equation*}
where $Y = \ln(1/x)$, $\tau = r Q_s(x)$, and the factorization scale given by $Q_s(x) = Q_0 \left( \frac{x_0}{x}\right)^{\lambda /2}$,  
\begin{eqnarray*}
a &=&-\frac{\ln (1-{N_0})}{\ln ^2 \left[ (1-{N_0})^{\frac{1}{\gamma_s
   }-\frac{1}{{N_0} \gamma_s }}\right]}\\
b &=& \frac{1}{2} (1-N_0)^{\frac{1}{\gamma_s }-\frac{1}{N_0 \gamma_s }}
\end{eqnarray*} 
and 
\begin{equation*}
\gamma_{eff}(x,r) = \gamma_s + \frac{\ln (2/\tau)}{\kappa \lambda Y}~.
\end{equation*} 

The Soyez dipole parameter values \cite{Soyez:2007kg} are $N_0 = 0.7 $, $\gamma_s = 0.738$, $\lambda = 0.220$, $x_0 = 0.163 \times 10^{-4}$, $\kappa = 9.94$, and $\sigma_0 = 27.3\ {\rm mb}$. The Munier dipole parameters  \cite{Munier-note}
shown in Fig. \ref{fig:dipole-comparison} 
are $N_0 = 0.7 $, $\gamma_s = 0.627$, $\lambda = 0.175$, $x_0 = 0.19 \times 10^{-6}$, $\kappa = 9.94$, and $\sigma_0 = 37.5\ {\rm mb}$.

\bibliographystyle{apsrev}
\bibliography{xsection-plus}

\end{document}